\documentclass[useAMS,usenatbib]{mn2e}
\usepackage{graphicx,float,amssymb,bm,xcolor,amsmath,wasysym}
\title[Type Ib Supernova iPTF13bvn]{Optical observations of the fast declining type Ib supernova iPTF13bvn}

\author[Srivastav et al.]{Shubham Srivastav\thanks{E-mail : ssrivastav@iiap.res.in}, 
G.~C.\ Anupama\thanks{E-mail : gca@iiap.res.in}, D.~K.\ Sahu\thanks{E-mail : dks@iiap.res.in}\\
Indian Institute of Astrophysics, Koramangala, Bangalore 560 034, India}

\begin{document}

\maketitle

\begin{abstract}
 We present optical UBVRI photometry and medium resolution spectroscopy of the type Ib supernova iPTF13bvn, 
 spanning a phase of $\sim$ $-13\,$d to $+71\,$d with respect to $B$-band maximum. The post maximum decline rates indicate 
 a fast decline with $\Delta m_{15}(B) = 1.82$. 
 Correcting for a galactic extinction $E(B-V){\rm_{MW}}=0.045$ and host galaxy extinction of $E(B-V){\rm_{host}}=0.17$,
 the absolute $V$-band magnitude peaks at M$_V=-17.23\, \pm \,0.20$. 
The bolometric light curve indicates that $\sim 0.09$ M$_{\odot}$ of $^{56}$Ni was synthesized in the explosion. 
The earliest spectrum ($-13$d) shows the presence of He~{\sc i} 5876 \AA\ feature at a velocity of $\sim$15000 km s$^{-1}$, 
which falls rapidly by the time the SN approaches the epoch of B-band maximum. 
The photospheric velocity near maximum light, as indicated by the Fe~{\sc ii} 5169~\AA\ feature, is $\sim 9000$ km s$^{-1}$.
The estimate for the $^{56}$Ni mass, together with the estimates for the ejected mass ($M_{\rm{ej}}$) and kinetic energy 
of the explosion ($E_{\rm{k}}$) indicate that iPTF13bvn is a low luminosity type Ib supernova, with a lower than average 
ejected mass and kinetic energy. Our results suggest that the progenitor of iPTF13bvn is inconsistent with a single 
Wolf-Rayet star.
\end{abstract}


\section{Introduction}
\label{sec:intro}

Core-collapse Supernovae (CCSNe) are caused by the violent deaths of massive ($>$ 8M$_{\odot}$) stars which have exhausted 
their nuclear fuel. Type Ib supernovae (SNe Ib) are a sub-class of the hydrogen deficient Stripped Envelope CCSNe, 
which are identified by the presence of optical He~{\sc I} lines in their spectra (see \citealt{Filippenko1997} for a review). 
The two plausible progenitor scenarios involve either a massive Wolf-Rayet star \citep{Gaskell1986} which has lost 
most of its outer envelope either through mass transfer to a companion or through strong stellar winds; 
or a relatively lower mass progenitor in a close binary system, eg. \citet{Podsiadlowski1992}; \citet{Nomoto1995}; see
\citet{Smartt2009} for a review.
SNe Ibc, and CCSNe in general show a rich diversity in their photometric and spectroscopic characteristics, unlike the 
fairly homogeneous sub-class of SNe Ia. The diversity in the observed properties of CCSNe is attributed to the diversity in
the nature and properties of the progenitor, such as its mass, radius, metallicity, mass-loss rate, rotation etc.
\newline
SN 1987A was the first supernova whose progenitor was unambiguously identified in pre-explosion images as a blue
supergiant \citep{Gilmozzi1987, Kirshner1987}.
Since then, several progenitors of type II-P SNe, the most common type of CCSNe, have been identified in pre-explosion 
images. However, a lot remains to be understood about the progenitors of Stripped Envelope CCSNe. \newline
The discovery of iPTF13bvn was reported by the intermediate Palomar Transient Factory (iPTF) on June $16.24$ UT in the 
host galaxy of NGC 5806 at a redshift of 0.005 \citep{atel5137}. The discovery magnitude of the transient was $18.0$ in 
the R-band, at the position RA = $15h \, 00m \, 00s.18$, Dec = $+01^\circ \, 52' \, 53''.5$. 
There was nothing at the position of the supernova on June $15.29$ UT, up to a limiting magnitude of $21.0$. 
Subsequently, it was classified as a young SN Ib on June 17.8 UT \citep{atel5142}. \newline
\citet{Cao13} present early phase photometry and spectroscopy of iPTF13bvn and also report a possible progenitor 
identification within a 2$\sigma$ error radius of 8.7 pc using HST pre-explosion images. The absolute luminosity, color
and inferred mass loss rate from radio data for this potential progenitor was found to be consistent with a single 
Wolf Rayet (WR) progenitor. Using stellar evolution models, \citet{Groh13} conclude that the properties of the candidate 
progenitor are consistent with that of a single WR star, in agreement with \citet{Cao13}. 
However, based on early and late phase observations of iPT13bvn and hydrodynamic modeling of the bolometric light curve, 
\citet{Fremling14} (hereafter F14) conclude that the bolometric light curve is inconsistent with a single WR star progenitor,
as previously suggested by \citet{Cao13} and Groh et al. (2013).
Further, \citet{Bersten14} (hereafter B14) perform hydrodynamic modeling of the bolometric light curve and propose an 
interacting binary system as the progenitor for iPTF13bvn. B14 also predict that the remaining companion, which is likely to
have O-type characteristics, could possibly be detected in the future with deep HST imaging once the supernova has 
faded sufficiently. Such a detection will provide the first confirmation for the interacting binary progenitor 
scenario for SNe Ibc.\\
In this paper, we present early phase optical photometry and spectroscopy of iPTF13bvn, followed by discussions based on the 
observed characteristics of the supernova.

\section{Data Reduction}

\subsection{Photometry}
\begin{figure}
\centering
\resizebox{0.85\hsize}{!}{\includegraphics{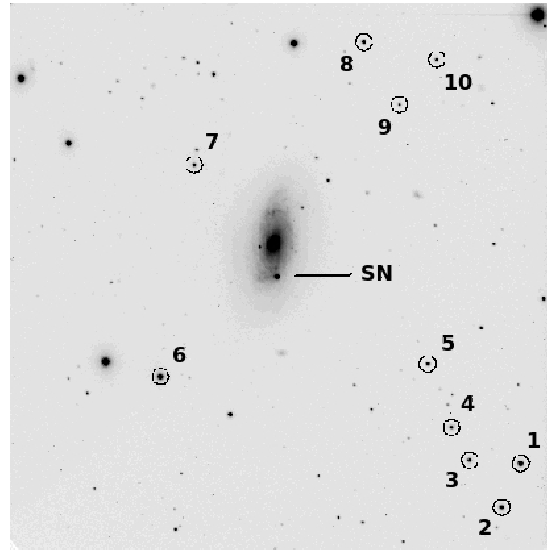}}
 \caption[]{\footnotesize Identification chart for iPTF13bvn. North is up and East is to the left. The field of view is
 $10' \, \times \, 10'$. Local standards are marked.}
 \label{fig:idchart}
\end{figure}

Photometric observations of iPTF13bvn commenced on 2013 June 18, $\sim$13 days before B-band maximum, using the 
2-m Himalayan Chandra Telescope at the Indian Astronomical Observatory at Hanle, and continued till 2013 September 10.
The observations were made with the Himalayan Faint Object Spectrograph Camera (HFOSC). The SITe CCD available
with the HFOSC has an imaging area of 2K $\times$ 4K pixels, of which the central unvignetted 2K $\times$ 2K area was used
for imaging. The field of view in imaging mode is 10 $\times$ 10 arcmin$^2$, with an image scale of 0.296 
arcsec pixel$^{-1}$. The supernova was imaged in the Bessell $UBVRI$ filters available with the HFOSC.
Landolt standard fields PG1633+099 and PG2213$-$006 \citep{Landolt1992} were observed under photometric conditions on 
the nights of 2013 July 13 and July 29 for photometric calibration of the supernova field.\newline 
Data reduction was performed in the standard manner using various packages available with the Image Reduction and 
Analysis Facility (IRAF\footnote{IRAF is distributed by the National Optical Astronomy Observatories, which are operated by
the Association of Universities for Research in Astronomy, Inc., under cooperative agreement with the National Science 
Foundation}). 
Aperture photometry was performed on the standard stars at an optimal aperture, determined using
the aperture growth curve method. An aperture correction was applied between the optimal aperture and an aperture close to
the FWHM of the stellar profile. Average extinction coefficients for the site \citep{Stalin08} were used in order to
account for atmospheric extinction and average values of the color terms for the HFOSC system were used to arrive at the 
photometric solutions.
The solutions thus obtained were used to calibrate several local standards in the supernova field, observed on the same
night as the standard fields.  The local standards were thereafter used to estimate the supernova magnitudes.
The magnitudes of the supernova and the secondary standards were evaluated using point-spread function (PSF) fitting, with
a fitting radius close to the full width at half-maximum (FWHM) of the stellar profile. 
Nightly photometric zero points were estimated using the local secondary standards and the supernova magnitudes 
were evaluated differentially. \newline
The field for iPTF13bvn is shown in Figure~\ref{fig:idchart}, and the magnitudes of the secondary standards are listed in 
Table~\ref{tab:secstd}. Table~\ref{tab:mag} summarizes the photometric observations and magnitudes of iPTF13bvn.

\begin{table*}
\centering
\caption{Magnitudes of secondary standards in the field of iPTF13bvn}
\label{tab:secstd}
\vspace{3mm} 
\begin{tabular}{c c c c c c}
 \hline \hline
ID & U & B & V & R & I \\ 
\hline
 1 & 16.76 $\pm$ 0.03 & 16.25 $\pm$ 0.03 & 15.37 $\pm$ 0.03  & 14.83 $\pm$ 0.02 & 14.35 $\pm$ 0.01 \\
 2 & 18.26 $\pm$ 0.05  & 16.92 $\pm$ 0.03 & 15.63 $\pm$ 0.03  & 14.85 $\pm$ 0.02 & 14.18 $\pm$ 0.02 \\
 3 & 16.45 $\pm$ 0.03  & 16.62 $\pm$ 0.02 & 16.05 $\pm$ 0.02  & 15.67 $\pm$ 0.03 & 15.29 $\pm$ 0.02 \\
 4 & 16.95 $\pm$ 0.03  & 16.80 $\pm$ 0.02 & 16.09 $\pm$ 0.02 & 15.67 $\pm$ 0.02 & 15.28 $\pm$ 0.02 \\ 
 5 & 17.31 $\pm$ 0.04  & 17.42 $\pm$ 0.03 & 16.76 $\pm$ 0.02 & 16.33 $\pm$ 0.03 & 15.89 $\pm$ 0.03 \\
 6 & 15.10 $\pm$ 0.03  &                  &                   &                  &		    \\
 7 & 		       & 17.07 $\pm$ 0.03 & 16.14 $\pm$ 0.02  & 15.55 $\pm$ 0.02 & 15.04 $\pm$ 0.02 \\
 8 & 17.28 $\pm$ 0.03  &			 &		     &			&	    \\
 9 & 		       & 17.48 $\pm$ 0.04 & 16.88 $\pm$ 0.03  & 16.48 $\pm$ 0.03 & 16.06 $\pm$ 0.03 \\
10 & 17.49 $\pm$ 0.04 & 17.41 $\pm$ 0.03 & 16.68 $\pm$ 0.02 & 16.22 $\pm$ 0.03 & 15.76 $\pm$ 0.02 \\
\hline

\end{tabular}

\end{table*}

\begin{table*}
\centering

\caption{Optical photometry of iPTF13bvn}
\label{tab:mag}
\vspace{4mm}
\begin{tabular}{c c c c c c c c}
\hline \hline
Date & JD & Phase$^*$ & U & B & V & R & I \\
 & (245 6000+) & (days) & & & & & \\
\hline 
2013/06/18 & 462.14 & $-$13.10 &                  & 17.69 $\pm$ 0.02 & 16.93 $\pm$ 0.01 & 16.71 $\pm$ 0.01 & 16.58 $\pm$ 0.01 \\
2013/06/19 & 463.13 & $-$12.11 &                  & 17.36 $\pm$ 0.02 & 16.69 $\pm$ 0.01 & 16.39 $\pm$ 0.01 & 16.26 $\pm$ 0.02 \\
2013/06/21 & 465.05 & $-$10.19 &                  & 16.71 $\pm$ 0.02 & 16.19 $\pm$ 0.02 & 15.87 $\pm$ 0.03 & 15.84 $\pm$ 0.02 \\
2013/06/22 & 466.14 & $-$9.10  &                  & 16.45 $\pm$ 0.03 & 15.99 $\pm$ 0.02 & 15.70 $\pm$ 0.01 & 15.63 $\pm$ 0.01 \\
2013/06/25 & 469.22 & $-$6.02  & 15.66 $\pm$ 0.03 & 15.95 $\pm$ 0.01 & 15.56 $\pm$ 0.01 & 15.33 $\pm$ 0.01 & 15.25 $\pm$ 0.02 \\
2013/06/28 & 472.24 & $-$3.00  & 15.64 $\pm$ 0.03 & 15.79 $\pm$ 0.01 & 15.37 $\pm$ 0.02 & 15.09 $\pm$ 0.01 & 14.97 $\pm$ 0.01 \\
2013/06/30 & 474.12 & $-$1.12  & 15.69 $\pm$ 0.02 & 15.78 $\pm$ 0.02 & 15.24 $\pm$ 0.01 & 15.01 $\pm$ 0.01 & 14.91 $\pm$ 0.02 \\
2013/07/01 & 475.23 & $-$0.01  &                  & 15.77 $\pm$ 0.01 & 15.20 $\pm$ 0.01 & 14.93 $\pm$ 0.01 & 14.84 $\pm$ 0.01 \\
2013/07/03 & 477.18 & +1.94    &                  & 15.83 $\pm$ 0.02 & 15.20 $\pm$ 0.01 & 14.91 $\pm$ 0.02 & 14.78 $\pm$ 0.01 \\
2013/07/04 & 478.21 & +2.97    & 15.99 $\pm$ 0.03 & 15.90 $\pm$ 0.02 & 15.23 $\pm$ 0.01 & 14.89 $\pm$ 0.01 & 14.75 $\pm$ 0.01 \\
2013/07/10 & 484.27 & +9.03    &                  & 16.64 $\pm$ 0.01 & 15.62 $\pm$ 0.01 & 15.06 $\pm$ 0.04 & 14.89 $\pm$ 0.02 \\
2013/07/11 & 485.22 & +9.98    &                  & 16.82 $\pm$ 0.01 & 15.73 $\pm$ 0.01 & 15.23 $\pm$ 0.01 & 14.99 $\pm$ 0.01 \\
2013/07/12 & 486.16 & +10.92   &                  &                  & 15.87 $\pm$ 0.01 & 15.37 $\pm$ 0.01 & 15.07 $\pm$ 0.01 \\
2013/07/13 & 487.25 & +12.01   & 17.79 $\pm$ 0.02 & 17.20 $\pm$ 0.01 & 15.99 $\pm$ 0.01 & 15.47 $\pm$ 0.01 & 15.15 $\pm$ 0.01 \\
2013/07/20 & 494.18 & +18.94   &                  & 17.86 $\pm$ 0.02 & 16.57 $\pm$ 0.01 & 15.94 $\pm$ 0.01 & 15.53 $\pm$ 0.02 \\
2013/07/23 & 497.12 & +21.88   &                  & 18.00 $\pm$ 0.02 & 16.73 $\pm$ 0.01 & 16.12 $\pm$ 0.01 & 15.74 $\pm$ 0.02 \\
2013/07/29 & 503.16 & +27.92   & 18.80 $\pm$ 0.03 & 18.19 $\pm$ 0.01 & 16.97 $\pm$ 0.01 & 16.33 $\pm$ 0.01 & 15.89 $\pm$ 0.01 \\
2013/08/06 & 511.17 & +35.93   &                  & 18.22 $\pm$ 0.02 & 17.18 $\pm$ 0.02 & 16.55 $\pm$ 0.02 & 16.10 $\pm$ 0.03 \\
2013/09/10 & 546.08 & +70.84   &                  &                  & 17.82 $\pm$ 0.03 & 17.36 $\pm$ 0.03 &                  \\
\hline
$^*$ \footnotesize{time since B-band max}
   \end{tabular}  
 
\end{table*}  

\subsection{Spectroscopy}

Spectroscopic observations of iPTF13bvn were made on nine epochs during 2013 June 18 and 2013 Aug 06, using grisms 
Gr7 (3500-7800 \AA) and Gr8 (5200-9250 \AA) available with the HFOSC. The details of the spectroscopic observations are 
summarized in Table~\ref{tab:speclog}.
Lamp spectra of FeAr and FeNe were used for wavelength calibration. The spectra were extracted in the standard manner.
Night sky emission lines $\lambda5577$, $\lambda6300$ and $\lambda6363$ were used to check the wavelength calibration 
and spectra were shifted wherever necessary. Spectra of spectrophotometric standards were observed in order to deduce 
instrumental response curves for flux calibration. For those nights where standard star observations were not available,
the response curves obtained during nearby nights were used. The flux calibrated spectra in grisms Gr7 and Gr8 were combined 
with appropriate scaling to give a single spectrum, which was brought to an absolute flux scale using the broadband 
$UBVRI$ photometric magnitudes. The spectra were corrected for redshift of the host galaxy. The telluric lines have not 
been removed from the spectra.

\begin{table}
 \centering
 \caption{Log of spectroscopic observations of iPT13bvn}
 \label{tab:speclog}
 \vspace{3mm}
 \begin{tabular}{c c c c}
 \hline
 Date & JD & Phase & Range \\
 & 245 6000+ & (days) & (\AA) \\
 \hline
 2013/06/18 & 462.16 & $-$13.08 & 3500-7800; 5200-9250 \\
 2013/06/19 & 463.17 & $-$12.07 & 3500-7800; 5200-9250 \\
 2013/06/25 & 469.24 & $-$6.0   & 3500-7800; 5200-9250 \\
 2013/06/28 & 472.20 & $-$3.04  & 3500-7800; 5200-9250 \\
 2013/07/01 & 475.24 & +0.0     & 3500-7800; 5200-9250 \\
 2013/07/04 & 478.23 & +2.99    & 3500-7800; 5200-9250 \\
 2013/07/11 & 485.23 & +9.99    & 3500-7800; 5200-9250 \\
 2013/07/23 & 497.13 & +21.89   & 3500-7800; 5200-9250 \\
 2013/08/06 & 511.19 & +35.95   & 3500-7800; 5200-9250 \\
 \hline
 \end{tabular}

\end{table}

\begin{figure}
\hspace{-7mm}
\resizebox{1.1\hsize}{!}{\includegraphics{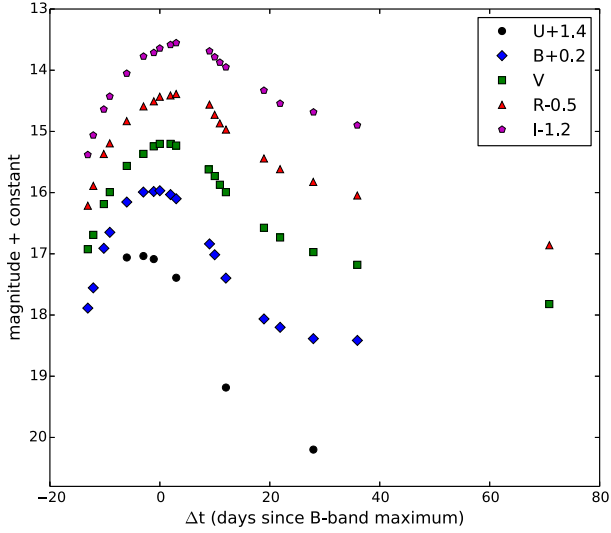}}
\caption[]{\footnotesize Optical light curves for iPT13bvn. The light curves have been shifted for clarity by the amount as 
indicated in the legend. The errors in the SN magnitudes are within or comparable to the size of the markers.}
\label{fig:lc}
\end{figure}

\section{Optical Light Curves}
\label{sec:olc}

iPTF13bvn was observed in Bessel $UBVRI$ filters for a period spanning a phase of $-13.1\,d$ to $+70.8\,d$ with respect to
$B$-band maximum. The epoch of maximum brightness and the peak magnitudes in each band were determined by fitting a 
spline curve through the points near maximum. The light curves in the bluer bands peak at earlier times. 
The supernova reached $B$-band maximum on JD 2456 475.24, at an observed apparent magnitude of 15.77 $\pm$ 0.02. 
The light curves are presented in Figure~\ref{fig:lc}.\newline
Using a power law model to fit the early $r$-band data for iPTF13bvn, \citet{Cao13} suggest the explosion date as June 
15.67 UT, which implies that the SN was detected within 1 day of the explosion. No signature of shock cooling was detected,
however, unlike in the case of SN 2008D \citep{Modjaz09}.
Using this explosion date for iPTF13bvn, the rise time to maximum in the $B$-band is found to be $\sim$16 days, 
whereas the rise time for the redder bands is longer, as seen in Table~\ref{tab:peak}. We use a rise time of 
16 days in this work.

The light curves show a fast decline, with $\Delta m_{15}(B)=1.82$.
The post-maximum decline rates are higher than those of most SNe Ibc in the literature, a notable exception being 
SN 1994I \citep{Richmond1996}. 
The light curves of iPTF13bvn are compared with a few well-studied SNe Ibc like
SN 2009jf \citep{Sahu11}, SN 2007gr \citep{Hunter09}, SN 2007Y \citep{Stritzinger09},
SN 1999ex \citep{Stritzinger02} and SN 1994I \citep{Richmond1996} in Figure~\ref{fig:lc_comp}. 
The reported magnitudes of SN 2007Y, which were in the $u$, $g$, $B$, $V$ and $r$ and $i$ bands, were transformed to the 
$UBVRI$ system using the transformations prescribed by \citet{Jester05}. 
The magnitudes of the supernovae have been normalized to their respective peak magnitudes and the time axis has been 
shifted to the epoch of maximum brightness in the $B$-band. The decline rates for iPTF13bvn suggest that the light curves 
are similar, though slightly faster declining than those of the narrow-lined type Ic SN 2007gr.

\begin{figure}
\hspace{-5mm}
\resizebox{\hsize}{!}{\includegraphics{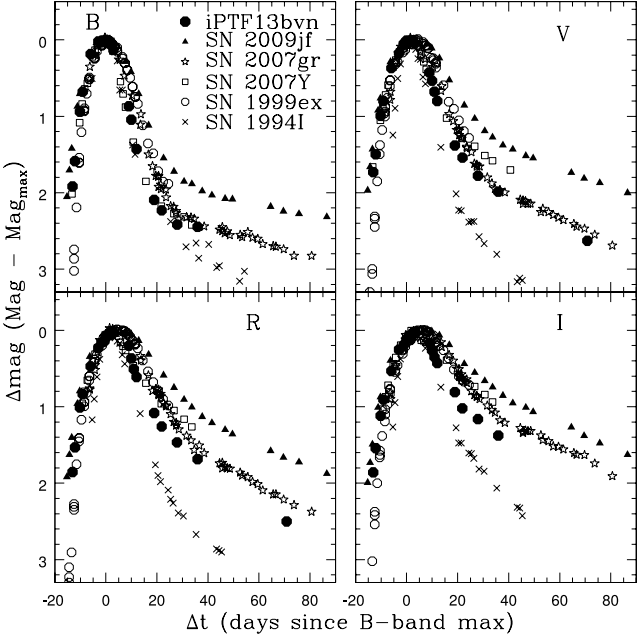}}
\caption[]{\footnotesize Comparison of BVRI light curves of iPTF13bvn, SN 2009jf, SN 2007gr, SN 2007Y, SN 1999ex and SN 1994I. 
The light curves have been shifted as described in the text.}
\label{fig:lc_comp}
\end{figure}

The color evolution of iPTF13bvn is shown in Figure~\ref{fig:color}.
The ($B-V$), ($V-R$) and ($R-I$) colors of iPT13bvn evolve towards blue in the pre-maximum phase. The ($B-V$) color reaches
a minimum at $\sim$5 days before B-band maximum.
Post maximum, the colors redden monotonically till $\sim$20 days past $B$-band maximum, beyond which the ($B-V$) color starts
becoming bluer again, whereas the ($V-R$) and ($R-I$) colors remain more or less constant.

\begin{figure}
\hspace{-5mm}
\resizebox{\hsize}{!}{\includegraphics{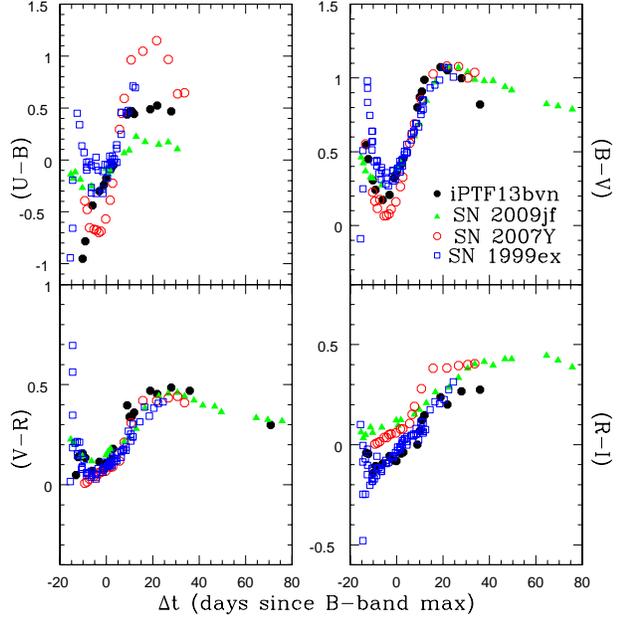}}
\caption[]{\footnotesize Color evolution of iPT13bvn, plotted along with that of SN 2009jf, SN 2007Y and SN 1999ex. 
The extinction correction applied was $E(B-V)_{\rm{MW}}=0.045$ and $E(B-V)_{\rm{host}}=0.17$, as discussed 
in Section ~\ref{sec:reddening}.}
\label{fig:lc_comp}
\end{figure}

\section{Spectral Evolution}
Spectroscopic observations of iPTF13bvn were obtained on 9 epochs, starting from $\sim$13 days before $B$-band maximum up to
$\sim$36 days past maximum.

\subsection{Pre-maximum Spectra}
\begin{figure}
\hspace{-6mm}
\resizebox{1.1\hsize}{!}{\includegraphics{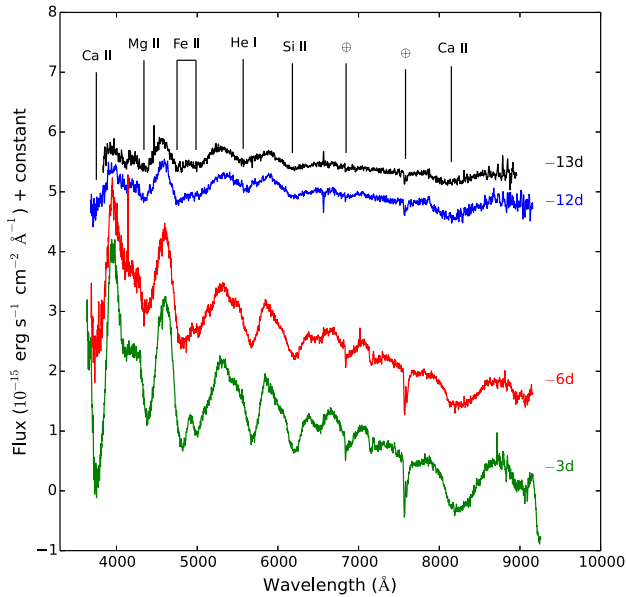}}
 \caption[]{\footnotesize Spectral evolution in the pre-maximum phase for iPTF13bvn. The prominent features in the
 spectra are marked.}
 \label{fig:premax}
\end{figure}

The pre-maximum spectra (Figure~\ref{fig:premax}) show broad P-Cygni profiles indicating a high expansion velocity of the
ejecta. The first spectrum, obtained 13 days before $B$-band maximum, shows the presence of He~{\sc i} 5876~{\AA} 
absorption feature at a velocity of $\sim$15000 km s$^{-1}$, with a possible contribution from Na~{\sc i} D 5890, 5896 \AA.
The He~{\sc i} 5876 \AA\ velocity was reported to be $\sim \, 16000 - 18000$ km s$^{-1}$ by \citet{atel5142} on 
June 17.8 UT, which corresponds to $\sim$14 days before $B$-band maximum. \newline
The spectra show an early emergence of He~{\sc i} 5876~{\AA} feature, similar to SN 2009jf \citep{Sahu11}. 
Other prominent features in the first spectrum are identified as Mg~{\sc ii} 4481 \AA, Fe~{\sc ii} 
lines (4555, 4924, 5018~{\AA}), Si~{\sc ii} 6355~{\AA}, O~{\sc i} 7774~{\AA} and the Ca~{\sc ii} NIR 
triplet (8498, 8542, 8662~{\AA}). The other features of He~{\sc i} at 4471, 5015, 6678, 7065~{\AA} are weak in the 
first spectrum but gradually become stronger as the supernova evolves towards maximum light. The He~{\sc i} 7065 \AA\ 
feature is affected by the telluric O$_2$ feature.
The continuum becomes bluer as the supernova approaches the epoch of $B$-band maximum, a trend also seen in the color 
evolution of the supernova. \newline
The absorption trough at $\sim$6200~{\AA} in the early spectra of SNe Ib is often attributed to a combination of
photospheric Si~{\sc ii} and/or high velocity ($\gtrsim 15000 \ {\rm km \ s}^{-1}$), detached H$\alpha$ 
\citep[eg.][]{Branch02, Anupama05, Folatelli06, Parrent07, Stritzinger09, Tanaka09}.
Using synthetic spectra generated with SYNOW, \citet{Elmhamdi06} argue that traces of Hydrogen are 
present in most, if not all SNe Ib spectra.
We identify the absorption trough at $\sim$6200~{\AA} in the early spectra of iPTF13bvn as Si~{\sc ii} 6355 \AA.
\citet{Cao13} suggest that the feature may also be associated with Ne I \citep[eg.][]{Benetti02}.

\begin{figure}
\hspace{-5mm}
\resizebox{1.1\hsize}{!}{\includegraphics{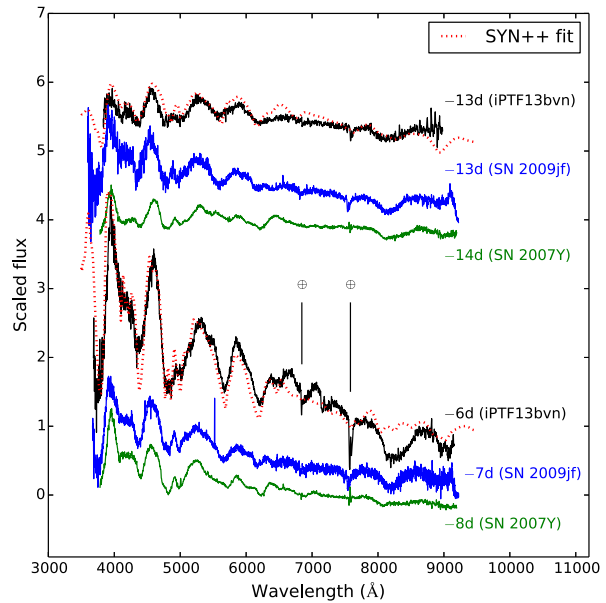}}
 \caption[]{\footnotesize Comparison of pre-maximum spectra of iPTF13bvn with SN 2009jf and SN 2007Y.}
 \label{fig:premax_comp}
\end{figure}

The pre-maximum spectra of SN iPTF13bvn are compared with those of SN 2009jf and SN 2007Y at similar epochs in 
Figure~\ref{fig:premax_comp}. iPTF13bvn shows remarkable similarity with SN 2009jf in the pre-maximum phase at similar
epochs, although the $-$6d spectrum of iPTF13bvn shows a bluer continuum when compared to the $-$7d spectrum of SN 2009jf. 
In contrast to iPTF13bvn and SN 2009jf, the He~{\sc i} and Ca~{\sc ii} NIR features in the early pre-maximum 
spectra of SN 2007Y are weak, but the Fe~{\sc ii} features present in the 4000--5000 \AA\ range are well developed.

\subsection{Post-maximum Spectra}

\begin{figure}
\hspace{-5mm}
\resizebox{1.1\hsize}{!}{\includegraphics{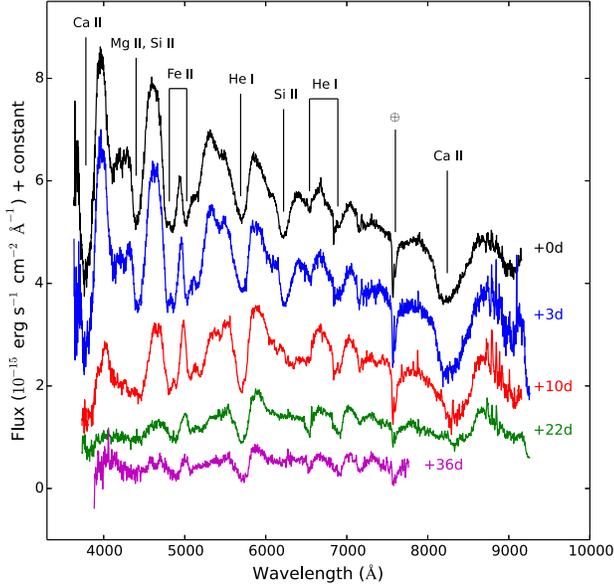}}
 \caption[]{\footnotesize Spectral evolution in the immediate post-maximum phase for iPTF13bvn.}
 \label{fig:postmax}
\end{figure}

The immediate post-maximum spectral evolution is shown in Figure~\ref{fig:postmax}. The spectrum obtained near maximum 
light exhibits a blue continuum and prominent signatures of He~{\sc i}, Fe~{\sc ii} and Ca~{\sc ii}. 
The He~{\sc i} features 5876, 6678 and 7065 \AA\ exhibit different expansion velocities and a single velocity in the 
synthetic spectrum is unable to fit the observed spectrum \citep[F14,][]{Cao13}. 
The absorption trough near 6200 \AA, which we associate with Si~{\sc ii} 6355 \AA\ weakens after maximum, and 
disappears beyond +10 days. A similar trend was seen for the Si~{\sc ii} 6355 \AA\ feature in the case of 
SN 2009jf \citep{Sahu11}.
The spectrum of iPTF13bvn obtained near maximum light is fit by a synthetic spectrum with photospheric velocity 
$v_{\rm{ph}}=9000$ km s$^{-1}$ and effective blackbody temperature $T_{\rm{bb}}=13000$ K, using SYN++. 
SYN++ is a modern C++ rewrite of SYNOW, the parameterized spectrum synthesis code \citep[see][]{Thomas11,Fisher2000}.
The spectrum obtained at +10 days is fit by a synthetic spectrum with reduced values of $v_{\rm{ph}}=6500$ km s$^{-1}$ and
$T_{\rm{bb}}=6000$ K, which further reduce to $v_{\rm{ph}}=5500$ km s$^{-1}$ and $T_{\rm{bb}}=5000$ K to fit the
+23 day spectrum.

\begin{figure}
 \hspace{-5mm}
\resizebox{1.1\hsize}{!}{\includegraphics{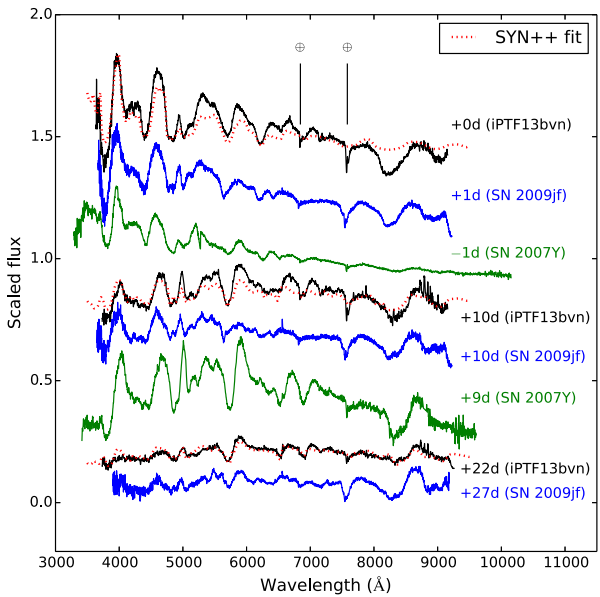}}
 \caption[]{\footnotesize Comparison of immediate post-maximum spectra of iPTF13bvn with SN 2009jf and SN 2007Y. The dotted 
 lines are the SYN++ fits to the spectra of iPTF13bvn.}
 \label{fig:postmax_comp}
\end{figure}

The maximum light and immediate post-maximum spectra of iPTF13bvn are compared with those of SN 2009jf and SN 2007Y at
similar epochs in Figure~\ref{fig:postmax_comp}. The maximum light spectrum of iPTF13bvn shows a good match with SN 2009jf.
At +10 days and beyond, the spectra of the three SNe look similar. 

\section{Ejecta expansion velocities}

The expansion velocities of various features were deduced by fitting a Gaussian to the minimum of the absorption trough in
the spectra corrected for the redshift of the host galaxy.
A synthetic spectrum generated using SYN++, with a photospheric velocity of $\sim$11000 km s$^{-1}$ deduced from 
Fe~{\sc ii} 5169 \AA\ feature, fits the first spectrum obtained at phase $-$13 days. The photospheric velocity
decreases steadily, reaching $\sim$9000 km s$^{-1}$ at the epoch of $B$-band 
maximum and dropping further to below $\sim$6000 km s$^{-1}$ around 20 days past maximum.
The He~{\sc i} 5876 \AA\ shows a velocity of $\sim$15000 km s$^{-1}$ at $-$13 days, which
drops down rapidly to $\sim$9500 km s$^{-1}$ near the epoch of maximum light, and further falls to $\sim$ 8300 km s$^{-1}$
at a phase of +36 days. The velocity evolution of the Ca~{\sc ii} NIR triplet follows that of the He~{\sc i} 5876 \AA\
feature, with a value of $\sim$14000 km s$^{-1}$ at $-$13 days but subsequently settles to a lower velocity of 
$\sim$7500 km s$^{-1}$ at +22 days.
The Si~{\sc ii} 6355 \AA\ feature has a relatively lower velocity of $\sim$7500 km s$^{-1}$ at $-$13 days, 
which falls to below 6000 km s$^{-1}$ post maximum. The lower velocity of
the Si~{\sc ii}, compared to the photospheric velocity is probably due to 
mixing. Modeling results of B14 indicate a high level of mixing. A similar 
trend in velocity evolution was also seen in the case of SN 2009jf 
\citep{Sahu11} and SN 2007Y \citep{Stritzinger09}.
The measured velocities for iPTF13bvn are consistent with those reported by \citet{Cao13} and F14.

\begin{figure}
\resizebox{1.1\hsize}{!}{\includegraphics{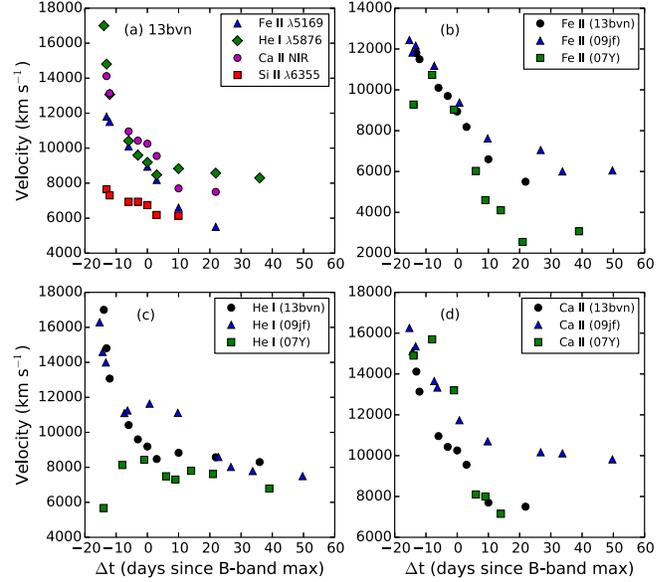}}
\caption[]{\footnotesize (a) Temporal evolution of velocity for the prominent features in the spectra of iPTF13bvn. 
(b), (c), (d) Comparison of Fe~{\sc ii} 5169~\AA, He~{\sc i} 5876 \AA\ and Ca~{\sc ii} NIR triplet velocities in
iPTF13bvn with SN 2009jf and SN 2007Y.}
\label{fig:velcomp}
\end{figure}

The velocity evolution of the He~{\sc i} 5876 \AA\ feature for iPTF13bvn matches well with SN 2009jf, whereas 
that of SN 2007Y exhibits a lower velocity of $\sim$8000 km s$^{-1}$ near maximum light. The evolution of the 
Fe~{\sc ii} 5169 \AA\ velocities in iPTF13bvn also matches with SN 2009jf till around maximum, beyond which it drops
to a lower value. 
The velocity evolution of prominent features in the spectra of
iPTF13bvn is shown in Figure~\ref{fig:velcomp}, along with a comparison of the velocity evolution of 
Fe~{\sc ii} 5169 \AA\, He~{\sc i} 5876 \AA\ and Ca~{\sc ii} NIR triplet with SN 2009jf \citep{Sahu11} and 
SN 2007Y \citep{Stritzinger09}.

\section{Distance and Reddening}\label{sec:reddening}

In order to correct for reddening, we adopt a galactic foreground extinction of $E(B-V)_{\rm{MW}}=0.045$ in the
direction of the host galaxy NGC 5806 \citep{S&F11}.
Since SNe Ibc are associated with dusty star-forming regions (\citealt{vanDyk1996}, \citealt{Anderson08}, \citealt{Kelly08}),
they typically suffer from significant host galaxy reddening. Using a sample consisting of 25 SNe Ibc, 
\citet{Drout11} find that the host galaxy reddening nearly always dominates the galactic foreground reddening, 
with an average host reddening of $E(B-V)_{\rm{host}}=0.21 \, \pm 0.20$. Combining their sample with those existing 
in the literature, Drout et al. estimate a mean host galaxy reddening of $E(B-V)_{\rm{host}}=0.36 \, \pm 0.24$ for SNe Ibc.\\
Comparing the ($B-V$) color of iPTF13bvn with an intrinsic color law derived from a sample of SNe Ibc, B14 estimate 
$E(B-V)_{\rm{host}}=0.17 \, \pm 0.03$. On the other hand, \citet{Cao13} provide a much lower value of 
$E(B-V)_{\rm{host}}=0.044$, deduced from the equivalent width of Na~{\sc i} D lines in high resolution spectra. 
This difference in the host reddening estimates is significant, corresponding to $\Delta A_V = 0.39$, considering a 
standard reddening law \citep{Cardelli1989} with $R_V=3.1$.

\citet{Drout11} suggest using the color evolution of SNe Ibc as an independent indicator for the line of sight 
reddening. Specifically, the scatter in the extinction-corrected ($V-R$) color curves is reported 
to be minimum at $\sim$10 days after maximum in $V$ and $R$ bands, with mean values of 
$\langle (V-R)_{V10} \rangle = 0.26 \, \pm \, 0.06$ and $\langle (V-R)_{R10} \rangle = 0.29 \, \pm \, 0.08$.
For iPTF13bvn, $(V-R)_{V10}=0.43$ and $(V-R)_{R10}=0.38$ for $E(B-V)_{\rm{host}}=0.044$; whereas  
$(V-R)_{V10}=0.34$ and $(V-R)_{R10}=0.29$ for $E(B-V)_{\rm{host}}=0.17$. In the latter case, the $(V-R)$ values of iPTF13bvn
are closer to the mean values suggested by \citet{Drout11}. Thus, the color evolution favors a host reddening
of $E(B-V){\rm_{host}}=0.17$. In the following analysis, we use $E(B-V)_{\rm{host}}=0.17$ for iPTF13bvn.

Adopting a distance modulus $\mu = 31.76 \, \pm \, 0.36$ \citep{Tully09}, which yields a distance of
$22.49^{+4.06}_{-3.44}$ Mpc, the estimated peak absolute magnitude in the $B$ and $V$ bands
is $-16.88\pm 0.20$ mag and $-17.25\pm 0.20$, respectively. Table~\ref{tab:peak} lists the estimated peak absolute
magnitudes in the $UBVRI$ bands. The errors in the absolute magnitudes were estimated taking into account the photometric 
errors on the observed magnitudes and the error in the distance modulus.
\begin{table*}
 \caption{Peak magnitudes, decline rates and colors for iPTF13bvn}
 \label{tab:peak}
 \vspace{3mm}
 \begin{tabular}{c c c c c c}
  \hline
Filter & Phase at peak & Decline Rate & Peak apparent magnitude & \multicolumn{2}{|c|}{Peak absolute magnitude} \\
       & (days) & $\Delta m_{15}$ & & $E(B-V)_{\rm{host}}=0.044$ & $E(B-V)_{\rm{host}}=0.17$\\
  \hline
  U & $-$3.77&                 & 15.63 $\pm$ 0.04 & -16.56 $\pm$ 0.19 & -17.17 $\pm$ 0.22\\ 
  B &  0.00  & 1.82 $\pm$ 0.05 & 15.77 $\pm$ 0.02 & -16.36 $\pm$ 0.17 & -16.88 $\pm$ 0.20\\
  V & +0.92  & 1.16 $\pm$ 0.04 & 15.19 $\pm$ 0.02 & -16.85 $\pm$ 0.17 & -17.25 $\pm$ 0.20\\
  R & +5.64  & 1.22 $\pm$ 0.04 & 14.83 $\pm$ 0.02 & -17.12 $\pm$ 0.17 & -17.44 $\pm$ 0.20\\
  I & +5.10  & 0.90 $\pm$ 0.04 & 14.72 $\pm$ 0.02 & -17.19 $\pm$ 0.17 & -17.38 $\pm$ 0.20\\
  \hline
  Color & $E(B-V)_{\rm{host}}=0.044$ & $E(B-V)_{\rm{host}}=0.17$ \\
 \hline
  $(U-B)_{Bmax}$ & -0.09 & -0.18 \\
  $(B-V)_{Bmax}$ & 0.48  & 0.35 \\
  $(V-R)_{Bmax}$ & 0.21  & 0.11 \\
  $(R-I)_{Bmax}$ & 0.02  & -0.08 \\
  $(V-R)_{V10}$  & 0.43  & 0.34 \\
  $(V-R)_{R10}$  & 0.38  & 0.29 \\
  \hline
 \end{tabular}
\end{table*}

\section{Bolometric Light Curve}
\label{sec:bol}

\begin{figure}
\hspace{-8mm}
\resizebox{1.1\hsize}{!}{\includegraphics{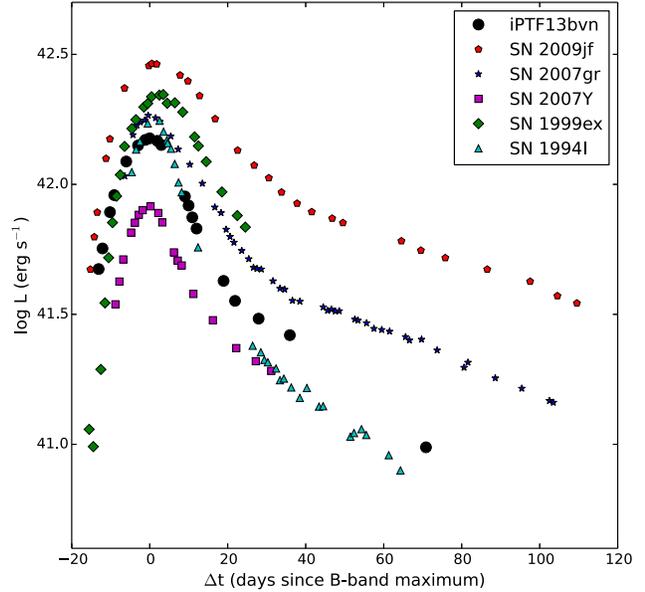}}
 \caption[]{\footnotesize Bolometric light curve of iPTF13bvn, plotted along with bolometric light curves of SN 2009jf, 
 SN 2007gr, SN 2007Y, SN 1999ex and SN 1994I for comparison.}
 \label{fig:bol_comp}
\end{figure}

The quasi-bolometric light curve for iPTF13bvn was constructed by converting the extinction corrected broadband $UBVRI$ 
magnitudes to monochromatic fluxes using the zero points from \citet{Bessell1998}. A spline curve was then fit through 
the monochromatic fluxes and the resulting curve was integrated under appropriate limits of wavelength as 
determined from the filter response curves, to yield the quasi-bolometric flux for the particular epoch.
Since $U$-band observations are not available beyond phase +28, we use only the $BVRI$ magnitudes to estimate the bolometric
flux thereafter.
The quasi-bolometric light curve is plotted in Figure~\ref{fig:bol_comp}, along with the bolometric light curves of 
SN 2009jf, SN 2007gr, SN 2007Y, SN 1999ex and SN 1994I for comparison. 
The quasi-bolometric light curves of SN 2009jf \citep{Sahu11}, SN 2007gr \citep{Hunter09}, SN 2007Y \citep{Stritzinger09},
SN 1999ex \citep{Stritzinger02} and SN 1994I \citep{Richmond1996} were constructed in a similar manner using the published
$UBVRI$ magnitudes. A total reddening $E(B-V)$ of 0.112, 0.092, 0.11, 0.3 and 0.45 was adopted and distances were
assumed to be 34.66, 29.84, 19.31, 48.31 and 8.32 Mpc for SN 2009jf, SN 2007gr, SN 2007Y, SN 1999ex and SN 1994I, 
respectively.\\

Assuming the date of explosion to be June 15.67 UT \citep{Cao13}, the bolometric light curve rises to a peak in
$\sim$16 days. With a peak bolometric magnitude of $-16.76 \pm 0.20$ (for $E(B-V)_{\rm{host}}=0.17$), iPT13bvn is 
$\sim$0.8 mag brighter than the SN Ib 2007Y, $\sim$0.20 mag fainter than SNe Ic 1994I and 2007gr, and $\sim$0.7 mag 
fainter than the normal SN Ib 2009jf.
The radioactive tail of the bolometric light curve (past 60 days of explosion) shows a fast decline rate of 
3 mag 100 d$^{-1}$, which is much higher than the theoretically expected rate of 0.98 mag 100 d$^{-1}$ powered by the
decay of radioactive $^{56}$Co. The fast decline indicates an optically thin ejecta with inefficient $\gamma$-ray trapping at 
later phases.\\

We note that the contribution of the Ultraviolet (UV) and Near Infrared (NIR) bands has not been included in the 
calculation of the bolometric light curve. For SN 2007Y, \citet{Stritzinger09} found a UV contribution of 
$\sim$20 \% and NIR contribution of $\sim$5 \% at maximum light. However, by two weeks past maximum, the UV flux 
drops to $\leq$10 \%, while the NIR contribution rises to $\sim$20 \% of the total bolometric flux. 
\citet{Modjaz09} report a NIR contribution of $\sim$20 \% for SN 2008D near two weeks past maximum.
Thus, the UV and NIR bands together can contribute $\sim$30 \% of the bolometric flux at any given epoch.

\section{Physical Parameters of the Explosion}

A combination of three parameters, namely the mass of $^{56}$Ni synthesized in the explosion ($M{\rm_{Ni}}$), 
the total ejecta mass ($M{\rm_{ej}}$) and the kinetic energy of the explosion ($E{\rm_{k}}$) determines the peak luminosity
and the width of the bolometric light curve \citep{Arnett1982}. A higher peak luminosity indicates a larger value of 
$M{\rm_{Ni}}$, and a larger ejecta mass results in broader light curves \citep{Ensman&Woosley1988}. 
The diffusion time $\tau_{m}$, which is a measure of the width of the bolometric light curve, is related to 
$M{\rm_{ej}}$ and $E{\rm_{k}}$ as $\tau_{m} \propto \kappa^{1/2} M{\rm_{ej}}^{3/4} E{\rm_{k}}^{-1/4}$, whereas the 
photospheric expansion velocity of the ejecta is related to $M{\rm_{ej}}$ and $E_{\rm{k}}$ as 
$v_{\rm{ph}} \propto \sqrt{E_{\rm{k}} / M_{\rm{ej}}}$ 
\citep[see][]{Arnett1982}. \newline
Arnett's rule suggests that the mass of $^{56}$Ni synthesized in the explosion can be estimated 
by equating the peak bolometric luminosity with the instantaneous rate of radioactive decay. \citet{Nugent1995} proposed a 
simplified formulation of Arnett's rule to calculate the $^{56}$Ni mass,
$$M_{\rm{Ni}} = \frac{L_{\rm{bol}}^{\rm{max}}}{\alpha\dot{S}}$$
Here, $\alpha$ is the ratio of bolometric and radioactivity luminosity, whereas $\dot{S}$ refers to radioactivity luminosity
per unit nickel mass, which depends on the rise time to maximum. 
The quasi-bolometric luminosity of iPT13bvn peaks at $1.05^{+0.4}_{-0.4} \times 10^{42}$ ergs s$^{-1}$ 
(for $E(B-V){\rm_{host}}=0.044$) and $1.50^{+0.4}_{-0.4} \times 10^{42}$ ergs s$^{-1}$ (for $E(B-V){\rm_{host}}=0.17$). 
The uncertainties in the peak flux are primarily due to the uncertainty in the distance modulus. Assuming a rise time of
$16 \, \pm 1$ days (Section~\ref{sec:bol}), and $\alpha$ to be unity, the mass of $^{56}$Ni is estimated to be 
$0.05^{+0.02}_{-0.02}$ M$_{\odot}$ and $0.08^{+0.02}_{-0.02}$, for $E(B-V){\rm_{host}}$ values of 0.044 and 0.17,
respectively.

In order to estimate the physical parameters of the explosion, such as $M_{\rm{Ni}}$, $M_{\rm{ej}}$ and 
E$_{\rm{k}}$, we fit the observed quasi-bolometric light curve of iPTF13bvn with the simple analytical model proposed by 
\citet{Vinko04}. 
This model assumes homologous expansion of the ejecta, and a constant opacity for $\gamma$-rays and positrons. 
A core-shell density structure is assumed, with a constant density core of fractional radius $x_0$, and
a surrounding shell whose density decreases outward as a power law with exponent $n$. A short diffusion time is assumed, 
which allows the emitted bolometric luminosity to be approximated as the rate of energy deposition. 
This assumption is however not valid during pre-maximum epochs when the ejecta is optically thick, and therefore this 
model is applied only to the post-maximum phase of the bolometric light curve.
Following \citet{Vinko04}, the time-dependent rate of energy deposition by $\gamma$-rays and positrons can be 
expressed as $$\epsilon = E_{\gamma} \, (1-exp(-\tau_{\gamma})) + E_+ \, (1-exp(-\tau_+)),$$
where E$_{\gamma}$, E$_+$ are energy input rates from $\gamma$-rays and positrons, and $\tau_{\gamma}$, $\tau_+$ denote
the optical depths for $\gamma$-rays and positrons respectively \citep{Capellaro97}, which in turn can be expressed as 
$$\tau_{\gamma , +} = \kappa_{\gamma , +}\rho_0(t)R(t)x_0(1 + \frac{1-x_0^{n-1}}{n-1}).$$ 
Here, $\rho_0(t) = \frac{M_{\rm{ej}}}{4\pi R(t)^3f(x_0)}$ is the time-varying core density, 
where $R(t) = v_{\rm{exp}}t$ (homologous expansion) and $f(x_0$) is a geometrical factor due to the assumed density
configuration.
The kinetic energy of the explosion then becomes 
$$E_{\rm{k}} = \frac{3}{10} \, \frac{3-n}{5-n} \, \frac{5x_0^n - nx_0^5}{3x_0^n-nx_0^3} \, M_{\rm{ej}}v_{\rm{exp}}^2.$$
\citet{Vinko04} define three models based on the assumed density configuration characterized by $x_0$ and $n$. 
Model A is a constant density model with $n = 0$, Model B is a power-law model with $x_0$ = 0.01 and Model C is a 
core-shell model with $x_0$ = 0.15.
The free parameters in the models are ejected mass ($M_{\rm{ej}}$), nickel mass ($M_{\rm{Ni}}$), 
$\gamma$-ray opacity ($\kappa_{\gamma}$), positron opacity ($\kappa_{+}$), expansion velocity ($v_{\rm{exp}}$) and the 
density power law exponent ($n$).

\begin{figure}
\resizebox{\hsize}{!}{\includegraphics{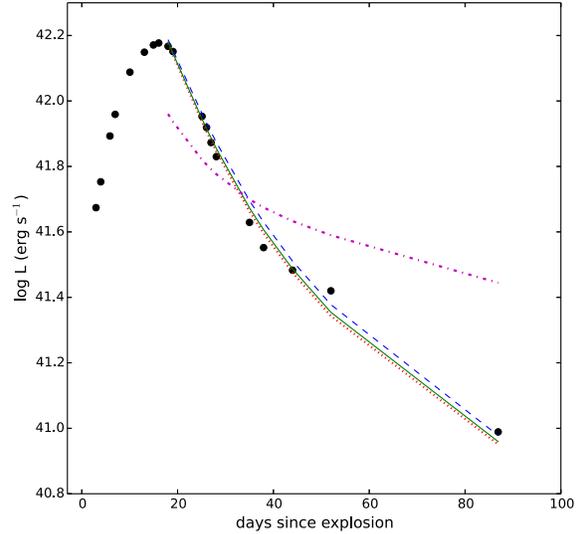}}
 \caption[]{\footnotesize Comparison of the observed bolometric light curve of iPTF13bvn with the computed models A, B, C 
 and C1.}
 \label{fig:modfits}
\end{figure}

Table ~\ref{tab:params} summarizes the best-fit values of the parameters and Figure~\ref{fig:modfits} shows the
results of the model fits to the bolometric light curve. 
The three models give similar results, with the models A and C fitting the observed data slightly better. The models 
favor a low value of the gamma-ray opacity ($\kappa_{\gamma} < 0.01$), in contrast to the usual value of 0.027 for grey
atmospheres \citep{Sutherland84}. The constant density model A yields slightly higher values of opacities, as also seen in
the case of SN 2002ap \citep{Vinko04}. The above models assume spherical symmetry and homologous
expansion of the ejecta. However, owing to the binary nature of the progenitor (B14), the ejecta is likely to be aspherical.
Therefore, the results tabulated in Table~\ref{tab:params} should only be treated as an order-of-magnitude estimate of the 
parameters, as also noted by \citet{Vinko04}.
Defining an additional model C1 where the value of $\kappa_{\gamma}$ is fixed at 0.027, the fit is found to be 
poorer (Fig~\ref{fig:modfits}). Fixing the $\gamma$-ray opacity in model C1 results in
a model whose peak luminosity is too low, and late phase decline too slow to explain the observed bolometric light curve. 

Hydrodynamic modeling of the bolometric light curve of iPTF13bvn requires $^{56}$Ni to be highly mixed in 
the outer layers of the ejecta (B14, F14). If the $^{56}$Ni is highly mixed, $\gamma$-ray escape commences earlier and the 
bolometric light curve is fainter and declines faster during the radioactive tail \citep{Woosley94}.
The He~{\sc i} features observed in SNe Ib spectra require nonthermal excitation caused by $\gamma$-rays emitted by the
decay of $^{56}$Ni freshly synthesized in the explosion \citep{Lucy91}.
The early emergence of the He~{\sc i} features in the spectra of iPTF13bvn indicates either a thoroughly mixed ejecta, 
or a small ejecta mass which would facilitate escape of $\gamma$-rays from the $^{56}$Ni-rich region, or both.
Thus, a low ejecta mass and a significant amount of mixing of $^{56}$Ni can explain the fast decline of the late phase 
bolometric light curve and the preference for models with low $\gamma$-ray opacity.

\begin{table*}
 \centering
 \caption{Parameter values for the computed models fit to the observed bolometric light curve \citep{Vinko04}. The asterisk
 sign indicates fixed parameter values.}
 \label{tab:params}
 \vspace{3mm}
 \begin{tabular}{c c c c c c c c c}
 \hline
 Model & $M_{\rm{ej}}$ & $M_{\rm{Ni}}$ & $E_{\rm{k}}$ & $\kappa_{\gamma}$ & $\kappa_+$ & $v_{\rm{exp}}$ & $x_0$ & $n$\\
       & (M$_{\odot}$) & (M$_{\odot}$) & ($10^{51}$ erg) & (cm$^2$g$^{-1}$) & (cm$^2$g$^{-1}$) & (km s$^{-1}$) & &\\
 \hline
 A     & 2.2 & 0.09 & 1.16 & 0.008 & 2.0 & 12000 & 1.0* & 0.0* \\
 B     & 1.5 & 0.09 & 1.25 & 0.003 & 1.0 & 14000 & 0.01*& 1.2 \\
 C     & 1.6 & 0.09 & 0.93 & 0.005 & 1.3 & 12000 & 0.15*& 1.2 \\
 C1    & 2.4 & 0.05 & 0.48 & 0.027*& 0.2 & 9700  & 0.15*& 3.1 \\
 \hline
 \end{tabular}

\end{table*}

We derive the physical parameters of the explosion as $M_{\rm{ej}} \sim 1.5 - 2.2$ M$_{\odot}$, 
$M_{\rm{Ni}} \sim 0.09$ M$_{\odot}$, $E_{51} \sim 1.0$ erg (Table~\ref{tab:params}). 
Accounting for a 30 \% (constant) contribution from missing passbands (UV and NIR) slightly raises these estimates to 
$M_{\rm{ej}} \sim 2 - 3$ M$_{\odot}$ and $ M_{\rm{Ni}} \sim 0.12$ M$_{\odot}$.
These estimates are consistent with those of F14 ($M_{\rm{ej}} \approx 1.94$ M$_{\odot}$, 
$M_{\rm{Ni}} \approx 0.05$ M$_{\odot}$, $E_{51} \approx 0.85$ erg) and B14 ($M_{\rm{ej}} \approx 2.3$ M$_{\odot}$, 
$M_{\rm{Ni}} \approx 0.1$ M$_{\odot}$, $E_{51} \approx 0.7$ erg), derived from detailed hydrodynamic modeling of the 
bolometric light curve. F14 use a host extinction correction of $E(B-V)_{\rm{host}}=0.044$, and hence obtain a lower value
of $M_{\rm{Ni}}$.
The above estimates for iPTF13bvn are lower than the median values for the SN Ib sample 
($M_{\rm{ej}} \approx 3.89$ M$_{\odot}$, $M_{\rm{Ni}} \approx 0.16$ M$_{\odot}$, $E_{51} \approx 2.3$ erg), 
as reported by \citet{Cano13}. Using the estimates provided by B14, the ratio $E_{\rm{51}}/M_{\rm{ej}}=0.3$ for iPTF13bvn, 
lower than the median value of $0.64 \pm 0.23$ for SNe Ib \citep{Cano13}. 

\begin{figure}
\resizebox{\hsize}{!}{\includegraphics{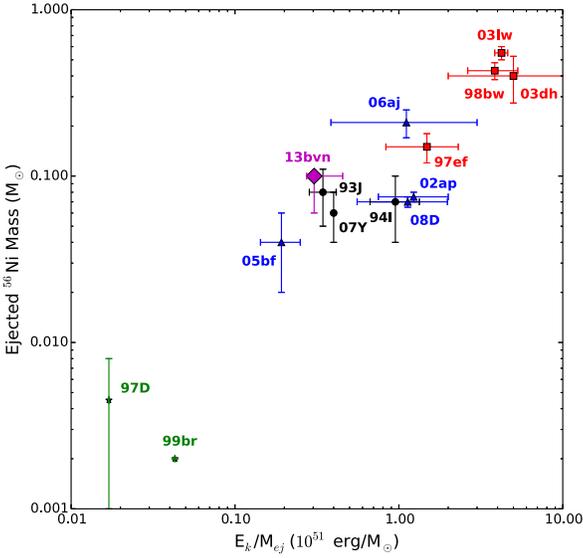}}
\caption[]{\footnotesize $M_{\rm{Ni}}$ vs $E_{\rm{k}}/M_{\rm{ej}}$ diagram for iPTF13bvn ($\blacklozenge$), 
plotted along with a few literature CCSNe. The parameters for iPTF13bvn are taken from B14, with error bars to accommodate 
the range of estimates provided in F14 and this work. ($\blacksquare$) denotes
hypernovae associated with Gamma Ray Bursts 
(GRBs) with $E_{\rm{51}}\gtrsim 10$, $\blacktriangle$ denotes broad-lined SNe 
(which includes both GRB and non-GRB SNe), 
$\bullet$ represents normal SNe, and $\bm{\star}$ represents faint/dark SNe \citep{Nomoto06}.}
\label{fig:Ibcdiag}
\end{figure}

One way to compare the physical parameters of CCSNe is to plot the $^{56}$Ni mass ($M_{\rm{Ni}}$) and kinetic 
energy ($E_{\rm{k}}$) against the estimated main-sequence mass ($M_{\rm{MS}}$) of the progenitor \citep[see][]{Nomoto06}. 
The estimated progenitor mass for iPTF13bvn is $\sim$20 M$_{\odot}$ (B14), similar to most normal SNe Ibc. 
However, the estimated values of $M_{\rm{Ni}}$ and $E_{\rm{k}}$ are similar to those estimated for SNe 1993J and 1994I.
Taking a slightly different approach, we plot $M_{\rm{Ni}}$ against the ratio $E_{\rm{k}}/M_{\rm{ej}}$
\citep[eg.][]{Bufano12} for iPTF13bvn along with a few other CCSNe in the literature (Figure~\ref{fig:Ibcdiag}). Except for 
SN 2007Y \citep{Stritzinger09} and iPTF13bvn, the values of $M_{\rm{ej}}$, $M_{\rm{Ni}}$ and $E_{\rm{k}}$ 
were taken from \citet{Tanaka09} and references therein. In the $M_{\rm{Ni}}$ vs $E_{\rm{k}}/M_{\rm{ej}}$ 
diagram, iPTF13bvn lies close to the type IIb SN 1993J and type Ib SN 2007Y. 
Both SN 1993J \citep{Shigeyama94} and SN 2007Y \citep{Stritzinger09} were relatively low luminosity events with 
$^{56}$Ni masses of 0.08 M$_{\odot}$ and 0.06 M$_{\odot}$, respectively, and low estimated progenitor masses of 
$\sim$13 M$_{\odot}$. The progenitor of SN 1993J was part of a binary system 
\citep[][and references therein]{Shigeyama94,Woosley94} and underwent mass loss via mass transfer to its companion. \\

The explosion parameters $M_{\rm{ej}}$ and $E_{\rm{k}}$ arrived at by hydrodynamic modeling (B14, F14), and the 
estimates provided here are incompatible with a single, massive progenitor with zero age main sequence (ZAMS) mass of 32 
M$_{\odot}$ suggested by Groh et al. (2013). Such a massive star would produce a helium star of $\sim$8 M$_{\odot}$, which 
is too massive to explain the observed properties of the SN (B14). Using hydrodynamic calculations coupled with stellar 
evolutionary calculations, B14 conclude that the progenitor was a low mass helium star (with a pre-explosion mass of 
$\sim$3.5 M$_{\odot}$) in a binary system, with an initial configuration of 20 M$_{\odot}+19$ M$_{\odot}$ and an initial
orbit of 4.1 days. Our results favor this model.

\section{Summary}

The optical light curves of iPTF13bvn show a fast decline ($\Delta m_{15}(B)=1.82$), which indicates a relatively small
ejecta mass. 
The peak absolute $V$-band magnitude, $V_{\rm{max}}=-17.25$ (corrected for $E(B-V)_{\rm{host}}=0.17$) is fainter than the
mean  extinction-corrected $V$-band magnitude of $-17.9 \, \pm \, 0.9$ for the SN Ib sample as reported by \citet{Drout11}.
The peak bolometric flux indicates that $\sim$ 0.05 -- 0.08 M$_{\odot}$ of $^{56}$Ni was synthesized in 
the explosion, depending upon the adopted value of host reddening.
Accounting for a constant 30 \% contribution from the UV and NIR passbands, the $^{56}$Ni mass estimate goes up to 
0.06 -- 0.09 M$_{\odot}$.
\\ \\
The optical spectra of iPTF13bvn are marked by an early emergence of the He~{\sc i} 5876 \AA\ feature, identified
in the first spectrum taken on day $-$13. The other features of He~{\sc i} at 4471, 6678 and 7065 \AA\ are weak in the first 
spectrum but are clearly identified in the spectrum taken on day $-$6. The photospheric velocity deduced from the 
Fe~{\sc ii} 5169 \AA\ feature is $\sim$9000 km s$^{-1}$ near maximum light, being consistent with the average photospheric
velocity of $8000 \, \pm \, 2000$ km s$^{-1}$ for SNe Ibc \citep{Cano13}. The spectra match well with those of the
type Ib SN 2009jf, although the light curves are very dissimilar.
\\ \\
Fitting the bolometric light curve with a simple analytical model yields $M_{\rm{Ni}} \sim 0.1$ M$_{\odot}$, 
$M_{\rm{ej}} \sim 2$ M$_{\odot}$ and $E_{\rm{k}} \sim 10^{51}$ erg.
The narrow peak of the bolometric light curve, and the subsequent fast decline indicates a small ejecta mass, thus being
inconsistent with a single, massive Wolf-Rayet progenitor for iPTF13bvn, as pointed out by B14 and F14. 

\section*{Acknowledgements}

We thank the staff of IAO, Hanle and CREST, Hosakote, that made these 
observations possible. The facilities at IAO and CREST are operated by the 
Indian Institute of Astrophysics, Bangalore. We also thank all the HCT observers
who spared part of their observing time for the ToO observations. This work has 
made use of the NASA Astrophysics Data System and the NED which is operated by 
Jet Propulsion Laboratory, California Institute of Technology, under contract 
with the National Aeronautics and Space Administration. We thank the anonymous
referee for the very positive comments.
 
\bibliography{biblist}

\end{document}